\documentclass[prb,twocolumn,showpacs,amscd,amsmath,amssymb,verbatim]{revtex4}

\usepackage{graphicx}
\usepackage{textcomp}
\usepackage[OT1,T1]{fontenc}

\begin{document}

\title{Magnetic versus nonmagnetic doping effects on the magnetic
ordering in the Haldane chain compound PbNi${}_{2}$V${}_{2}$O${}_{8}$}
\author{Andrej Zorko}
\email{andrej.zorko@ijs.si}
\affiliation{Jozef \v{S}tefan Institute, Jamova 39, 1000 Ljubljana,
Slovenia}\author{Denis Ar\v{c}on}
\affiliation{Jozef \v{S}tefan Institute, Jamova 39, 1000 Ljubljana,
Slovenia}\author{Alexandros Lappas}
\affiliation{Institute of Electronic Structure and Laser, Foundation
for Research and Technology -- Hellas, P.O. Box 1527, 71110 Heraklion,
Crete, Greece}\author{Zvonko Jagli\v{c}i\'c}
\affiliation{Institute of Mathematics, Physics and Mechanics, Jadranska
19, 1000 Ljubljana, Slovenia}\label{I1}\label{I2}\label{I3}
\date{\today}

\begin{abstract}
A study of an impurity driven phase-transition into a magnetically
ordered state in the spin-liquid Haldane chain compound PbNi${}_{2}$V${}_{2}$O${}_{8}$
is presented. Both, macroscopic magnetization as well as {}\textsuperscript{51}V
nuclear magnetic resonance (NMR) measurements reveal that the spin
nature of dopants has a crucial role in determining the stability
of the induced long-range magnetic order. In the case of nonmagnetic
(Mg${}^{2+}$) doping on Ni${}^{2+}$ spin sites ($S=1$) a metamagnetic
transition is observed in relatively low magnetic fields. On the
other hand, the magnetic order in magnetically (Co${}^{2+}$) doped
compounds survives at much higher magnetic fields and temperatures,
which is attributed to a significant anisotropic impurity-host magnetic
interaction. The NMR measurements confirm the predicted staggered
nature of impurity-liberated spin degrees of freedom, which are
responsible for the magnetic ordering. In addition, differences
in the broadening of the NMR spectra and the increase of nuclear
spin-lattice relaxation in doped samples, indicate a diverse nature
of electron spin correlations in magnetically and nonmagnetically
doped samples, which begin developing at rather high temperatures
with respect to the antiferromagnetic phase transition.
\end{abstract}
\pacs{75.30.Kz, 75.40.-s, 76.60.-k}
\maketitle

\section{INTRODUCTION}\label{XRef-Section-42294347}

Highly correlated quantum spin systems exhibit a variety of intriguing
magnetic phenomena in reduced dimensions. Impurities introduced
to such systems have been in the past successfully employed to reveal
the magnetic character of host materials. The virtue of magnetic
doping and boundaries created by nonmagnetic dopants is to introduce
novel degrees of freedom. Study of electronic structure and electron
spin correlations in the vicinity of the impurity sites is essential
for understanding the quantum state of the host. This approach has
been, for instance, widely exploited in the quest of finding an
interconnection between magnetic and superconducting nature of high-{\itshape
T}${}_{c}$ cuprates.\cite{PanNature403,HudsonNature411} Similarly,
doping turned out to be an irreplaceable source of information about
the ground state in various other transition-metal oxides exhibiting
macroscopically coherent ground states, such as singlet ground states
in $S=1/2$ spin-Peierls chains,\cite{HasePRL71} spin ladders\cite{AzumaPRB55}
and Haldane integer-spin chains,\cite{HagiwaraPRL65} which all show
a spin gap in their excitation spectrum.

Common consequences of the nonmagnetic impurities embedded in low-dimensional
antiferromagnets include an enhancement of antiferromagnetic correlations
and an induction of staggered magnetic moments near the impurity
sites. These effects are manifested in the vicinity of vacancies
in one-dimensional\cite{EggertPRL75} and two-dimensional\cite{BulutPRL62}
antiferromagnetic Heisenberg lattices with gapless excitations,
as well as in spin-gap systems with robust nonmagnetic ground states
including spin-Peierls chains,\cite{MartinsPRL78,LaukampPRB57} spin
ladders\cite{MotomeJPSJ65} and Haldane chains.\cite{MiyashitaPRB48,SorensenPRB49}
For this reason, it was argued that the enhanced correlations at
short distances have a common origin independent of the long-distance
behavior of the spin correlations. The enhancement can be explained
within the so-called ``pruned'' resonating-valence-bond picture
to be a consequence of pruning the possible singlet configurations
for spins close to the vacancy sites.\cite{MartinsPRL78,LaukampPRB57}
The staggered magnetic moments have been recently observed also
experimentally in the case of nonmagnetic\cite{TedoldiPRL83,DasPRB69}
and magnetic\cite{DasPRB69} doping of the Haldane chain compound
Y${}_{2}$BaNiO${}_{5}$, where exponentially decaying staggered magnetization
was detected next to the impurities, with the staggered moments
only weakly dependent on the magnetic nature of the dopants.

A rather unexpected feature observed in different one-dimensional
spin-gap systems is the presence of long-range magnetic ordering
induced by impurities.\cite{HasePRL71,AzumaPRB55,OOsawaPRB66,UchiyamaPRL83}
The ordering is believed to originate from the impurity-induced
staggered moments. These moments are coupled along the chains with
an exponentially decaying staggered exchange,\cite{SorensenPRB49}
because the coupling is mediated by virtual excitations of the gaped
medium.\cite{MelinEPJB18} Obviously, interchain coupling is vital
for providing foundations for the three-dimensional magnetic ordering.\cite{InakagiJPSJ52}
There have been few attempts, recently published, trying to unify
the field of the impurity-induced magnetic ordering. For instance,
M\'elin et al.\cite{MelinEPJB18} highlighted the role of three characteristic
temperatures in doped spin-gap systems; namely, the temperature
below which spin correlations within chains begin to develop, the
typical energy of the staggered exchange, and the Stoner temperature
of the order of the interchain coupling, associated with the development
of three-dimensional magnetic correlations. The relative amplitudes
of the latter two have been argued to determine whether the impurity-induced
long-range ordering should be present. Although there is a great
deal of reports on long-range ordering as a consequence of doping
different spin-gap systems, a comprehensive theory capable of ``predicting''
the observed transition temperatures and the stability of the induced
long-range magnetically ordered states upon external perturbations
such as magnetic field, still does not seem to be within reach.
Moreover, the effect of the magnetic character of the dopants on
magnetic ordering is particularly vague and, therefore, implores
for further investigation.

The purpose of this paper is, therefore, to provide a further experimental
insight into the complex problem of the impurity-induced phenomena
regarding the magnetism of spin-gap systems. The emphasis is put
on the comparison between magnetic and nonmagnetic doping of the
Haldane-chain compound PbNi${}_{2}$V${}_{2}$O${}_{8}$, possessing
chains of Ni${}^{2+}$ spins ($S=1$), with a by far dominant nearest-neighbor
({\itshape nn}) intrachain exchange coupling $J=95$~K (in units
of {\itshape k}{\itshape ${}_{B}$}).\cite{UchiyamaPRL83} dc magnetization
measurements are used to explore the temperature as well as the
magnetic-field stability of the impurity-induced long-range magnetic
order. In addition, nuclear magnetic resonance (NMR) measurements
on {}\textsuperscript{51}V nuclei weakly coupled to the system of
localized electronic moments through a transferred hyperfine coupling,
are exploited as a local-probe technique, capable of detecting the
inhomogeneities as well as the dynamics in the electron spin system.
We show that the magnetic nature of the dopants plays a crucial
role in reducing quantum fluctuations in Haldane chains and thus
helps stabilizing the magnetic long-range order. Moreover, it dictates
the character of the electron spin correlations at low-temperatures.

\section{EXPERIMENTAL}

All the samples used in this investigation were prepared according
to the same solid-state reaction procedure.\cite{LappasPRB66} The
resulting polycrystalline powders can be addressed as solid solutions,
since the Mg${}^{2+}$ and Co${}^{2+}$ substitutions for Ni${}^{2+}$
ions are randomly distributed. Phase purity of the samples was verified
by powder x-ray and neutron diffractometry,\cite{LappasPRB66,MastorakiJSSC177,MastorakiAPA74}
which revealed that the effect of the impurities on the position
of all the ions within the unit cell is minor. The same samples
were used in our previous report presenting combined results of
electron spin resonance (ESR) and magnetization measurements in
low magnetic field.\cite{ZorkoESRCoMg}
\begin{figure}[ht]
\begin{center}
\includegraphics[width=7.2cm]{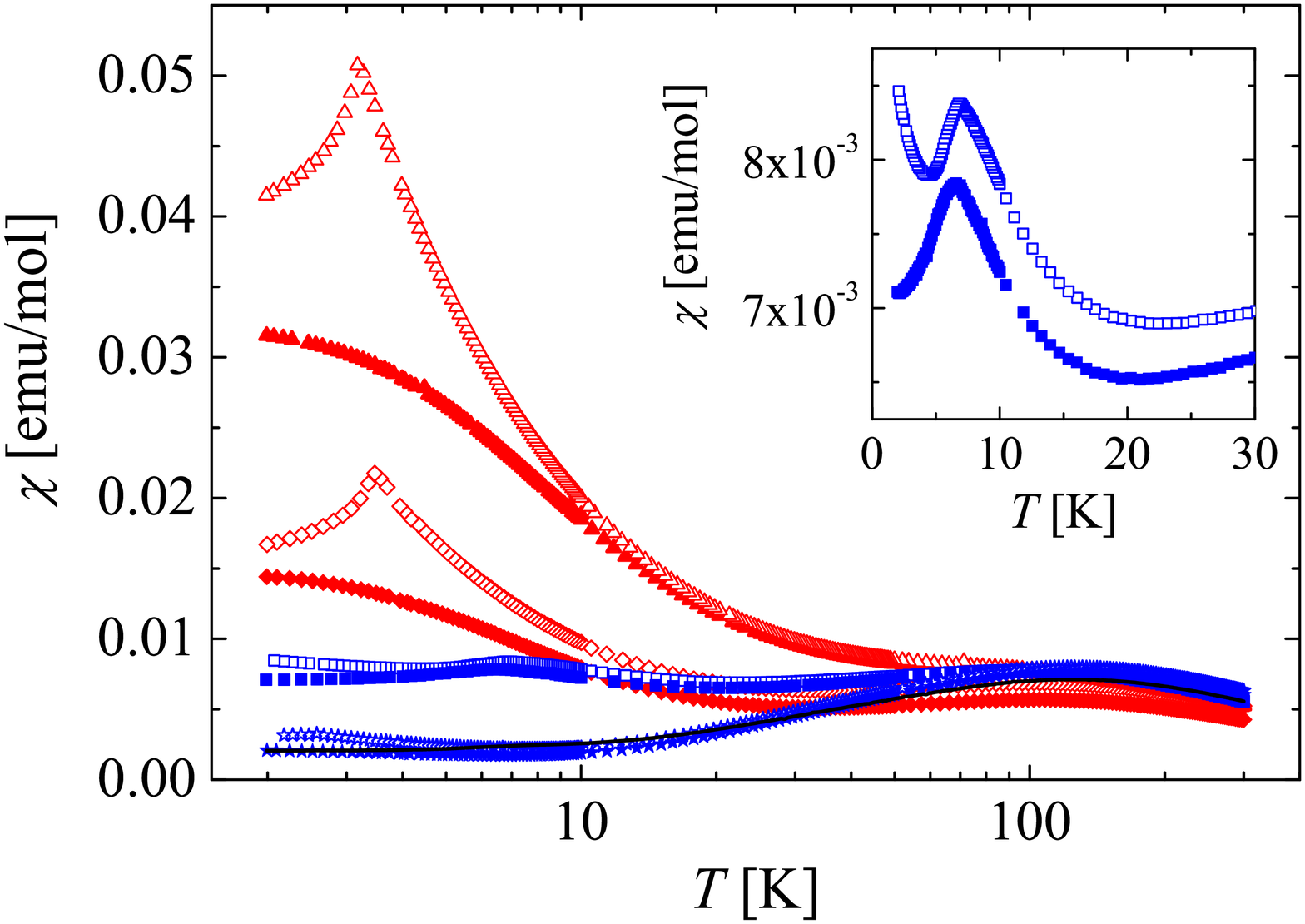}

\end{center}
\caption{ (Color online) dc susceptibility in PbNi${}_{2}$V${}_{2}$O${}_{8}$
(solid line), nonmagnetically doped PbNi${}_{1.88}$Mg${}_{0.12}$V${}_{2}$O${}_{8}$
(diamonds) and PbNi${}_{1.76}$Mg${}_{0.24}$V${}_{2}$O${}_{8}$ (triangles),
as well as magnetically doped PbNi${}_{1.98}$Co${}_{0.02}$V${}_{2}$O${}_{8}$
(stars) and PbNi${}_{1.92}$Co${}_{0.08}$V${}_{2}$O${}_{8}$ (squares)
in magnetic fields of 0.1~T (open symbols) and 5~T (full symbols
and solid line). The inset shows a decrease of the low-temperature
peak position with increased field in the PbNi${}_{1.92}$Co${}_{0.08}$V${}_{2}$O${}_{8}$
compound.}
\label{XRef-Figure-523213316}
\end{figure}

Bulk magnetic measurements were performed with a Quantum Design
SQUID magnetometer. dc magnetization was obtained from field-cooled
runs between room temperature and 2~K in static magnetic fields
of 0.1~T and 5~T.

{}\textsuperscript{51}V NMR measurements were conducted by standard
pulsed NMR techniques in an external magnetic field of 6.34~T in
the temperature range between room temperature and 4.2~K. Typical
width of applied $\pi$/2 pulses was 6 $\mu$s. Due to the rather
broad NMR spectra the spectral width of such pulses was insufficient
to reliably detect the full spectra even at room temperature. For
this reason, the NMR spectra were obtained in frequency-sweep experiments
from the amplitude of the Hahn-echo-detected signals. {}\textsuperscript{51}V
spin-lattice relaxation was investigated by employing a saturation-recovery
method. 

\section{RESULTS}

\subsection{dc magnetization}
\begin{figure}[ht]
\begin{center}
\includegraphics[width=7.2cm]{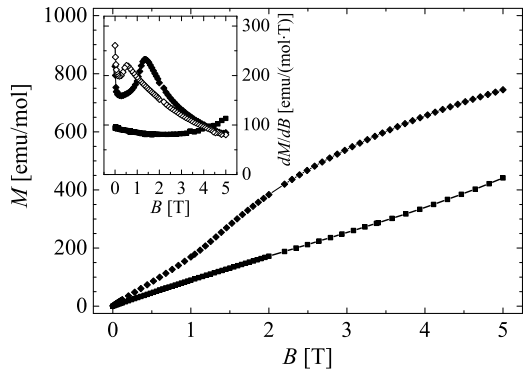}

\end{center}
\caption{ Magnetization curves for PbNi${}_{1.88}$Mg${}_{0.12}$V${}_{2}$O${}_{8}$
(diamonds) and PbNi${}_{1.92}$Co${}_{0.08}$V${}_{2}$O${}_{8}$ (squares)
measured at 2~K. Inset shows the derivative of both curves as well
as of the curve measured in the former compound at 3.2~K (open diamonds).}
\label{XRef-Figure-524151256}
\end{figure}

Upon doping, the PbNi${}_{2}$V${}_{2}$O${}_{8}$ compound undergoes
a phase transition from the spin-liquid Haldane state into the long-range
magnetically ordered state.\cite{UchiyamaPRL83} The antiferromagnetic
nature of the observed ordering was confirmed by additional Bragg
peaks in neutron diffraction patterns observed below the phase-transition
temperature.\cite{LappasPRB66} The ordering can also be detected
by dc magnetization measurements, which show clear peaks in dc susceptibility
at the phase-transition temperature. In Fig. \ref{XRef-Figure-523213316}
such measurements are presented for two Mg-doped compounds, PbNi${}_{1.88}$Mg${}_{0.12}$V${}_{2}$O${}_{8}$
and PbNi${}_{1.76}$Mg${}_{0.24}$V${}_{2}$O${}_{8}$, and two Co-doped
samples, PbNi${}_{1.98}$Co${}_{0.02}$V${}_{2}$O${}_{8}$ and PbNi${}_{1.92}$Co${}_{0.08}$V${}_{2}$O${}_{8}$.
In accordance with the previous measurements,\cite{UchinokuraPhysicaB,MastorakiAPA74}
the phase transition in an external magnetic field of 0.1~T is observed
in nonmagnetically doped samples PbNi${}_{1.88}$Mg${}_{0.12}$V${}_{2}$O${}_{8}$
and PbNi${}_{1.76}$Mg${}_{0.24}$V${}_{2}$O${}_{8}$ at 3.4~K and
3.2~K, respectively, while it is shifted to 7.1~K in the magnetically
doped PbNi${}_{1.92}$Co${}_{0.08}$V${}_{2}$O${}_{8}$ compound. The
transition in the lightly doped PbNi${}_{1.98}$Co${}_{0.02}$V${}_{2}$O${}_{8}$
compound is found around 2.6~K, similarly as reported elsewhere.\cite{ImaiCM2004}

The magnitude of the impurity-induced low-temperature dc susceptibility
substantially depends on the spin nature of the dopants as shown
in Fig. \ref{XRef-Figure-523213316}. This was successfully attributed
to the magnetic coupling between the impurity and the impurity-liberated
spins in our previous report.\cite{ZorkoESRCoMg} Based on the impurity-host
antiferromagnetic coupling in Co-doped samples, $J_{i-h}=14$~K,
and the effective ferromagnetic coupling between spins neighboring
a particular impurity site, ${\tilde{J}}^{\prime }\approx -2$~K,
both derived from the ESR investigations, we were able to give a
quantitative description of the observed phenomenon.

Upon increasing the static magnetic field up to 5~T, the peak in
the dc susceptibility disappears in the case of Mg-doping. On the
other hand, the peak is only slightly displaced to 6.5~K in the
PbNi${}_{1.92}$Co${}_{0.08}$V${}_{2}$O${}_{8}$ compound as shown
in the inset of Fig. \ref{XRef-Figure-523213316}. The latter observation
together with the significantly increased phase-transition temperature
in this sample indicate enhanced stability of the magnetic order
in the cobalt doped samples. The disappearance of the phase transition
in relatively low magnetic fields of the order of few Tesla has
already been reported for lightly Mg-doped compounds\cite{MasudaPRB66}
as well as for the heavily doped PbNi${}_{1.76}$Mg${}_{0.24}$V${}_{2}$O${}_{8}$
compound.\cite{LappasPRB66} In the latter case, a field-induced
metamagnetic transition was suggested based on the dc and ac susceptibility
results. Similar behavior is found also in the PbNi${}_{1.88}$Mg${}_{0.12}$V${}_{2}$O${}_{8}$
sample. More specifically, within the magnetically ordered phase
the field dependence of the magnetization curve exhibits a pronounced
inflection point (see Fig. \ref{XRef-Figure-524151256}), typical
of the second order phase transition from the antiferromagnetically
ordered to the paramagnetic phase.\cite{StryjewskiAP487} Such transition
can be recognized to occur around 1.4~T at 2~K, corresponding to
the peak in the derivative of the magnetization curve. The critical
field decreases with temperature. For instance, the peak is observed
around 0.5~T at 3.2~K (see Fig. \ref{XRef-Figure-524151256}).
\begin{figure}[t]
\begin{center}
\includegraphics[width=7.2cm]{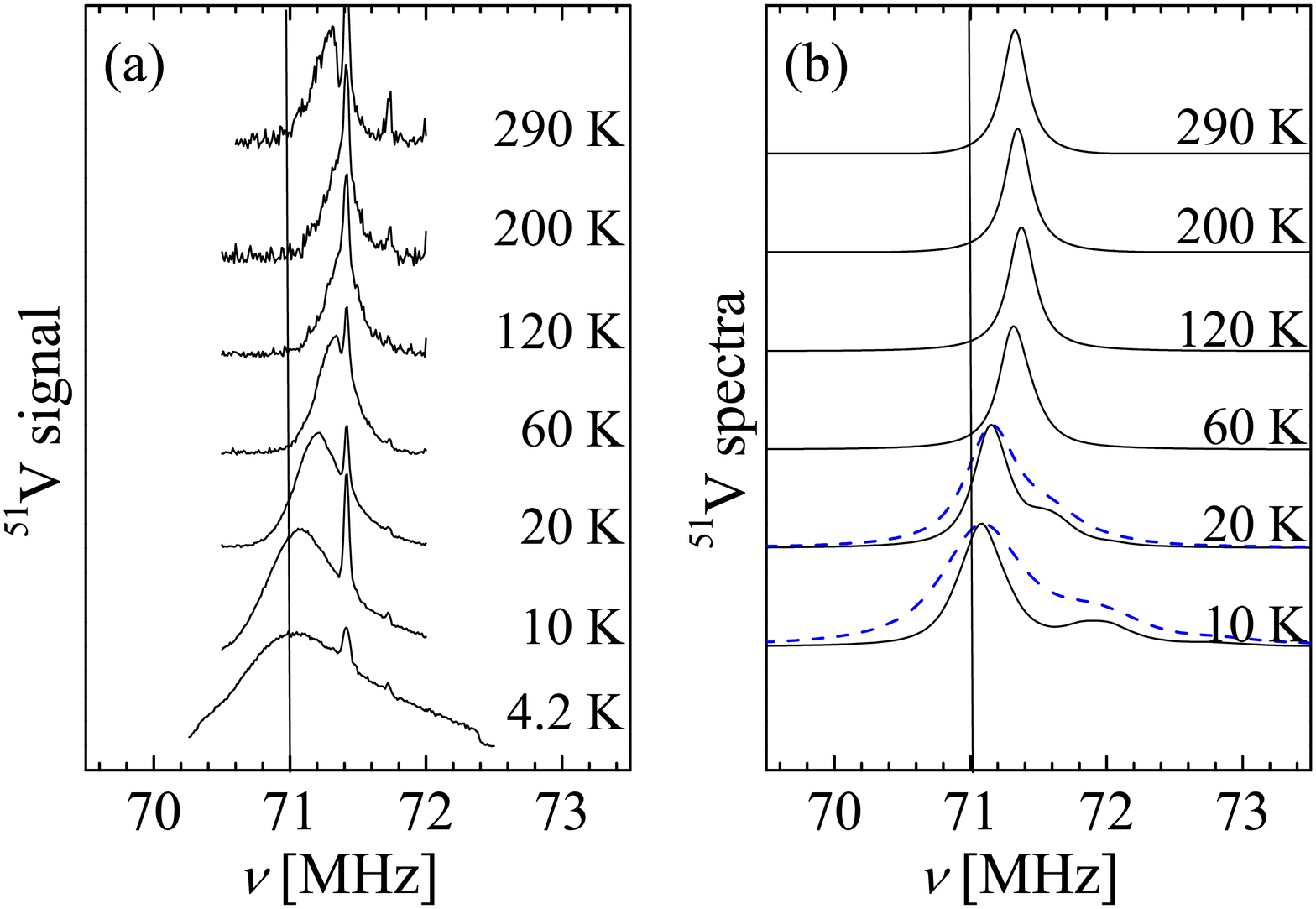}

\end{center}
\caption{ The collection of (a) measured and (b) simulated {}\textsuperscript{51}V
NMR spectra in PbNi${}_{1.88}$Mg${}_{0.12}$V${}_{2}$O${}_{8}$. Fits
represented by solid lines are based on the assumption of temperature
independent homogeneous broadening, while dashed lines correspond
to temperature dependent broadening.}
\label{XRef-Figure-523193630}
\end{figure}

On the contrary, in PbNi${}_{1.92}$Co${}_{0.08}$V${}_{2}$O${}_{8}$
a linear dependence of the magnetization curve is detected up to
4~T at 2~K (see Fig. \ref{XRef-Figure-524151256}). The curvature
of the magnetization curve is slightly enhanced above 4~T as expressed
in its derivative, however, it seems that even a magnetic field
of 5~T is still appreciably below the critical field value, which
would destroy the magnetic order. The antiferromagnetic order in
Co-doped samples obviously persists to much higher magnetic fields,
which is also in line with the results of the NMR measurements presented
below.

\subsection{NMR measurements}

\begin{figure}[ht]
\begin{center}
\includegraphics[width=7.2cm]{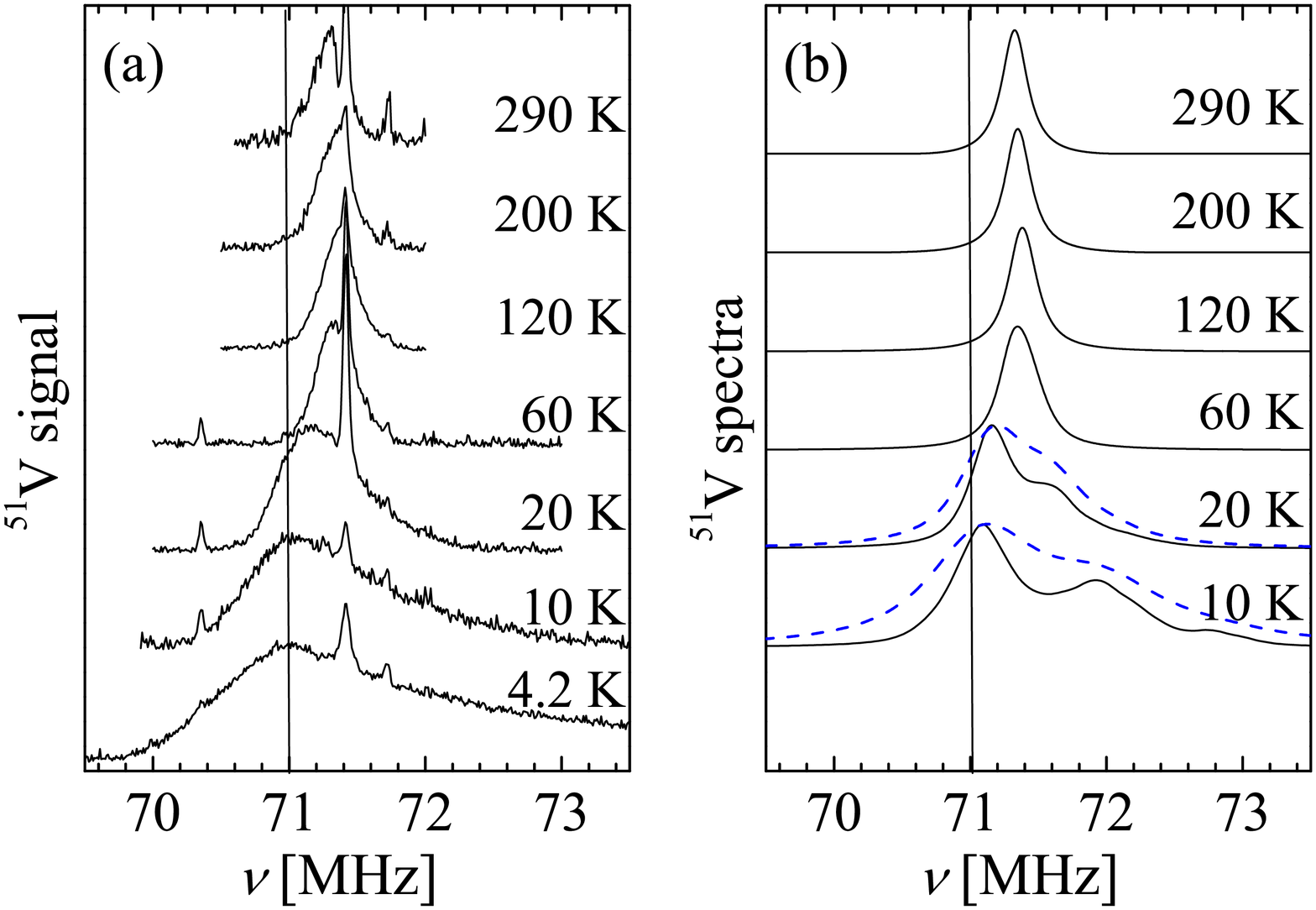}

\end{center}
\caption{ The {}\textsuperscript{51}V NMR spectra (a) measured and
(b) simulated for the case of PbNi${}_{1.76}$Mg${}_{0.24}$V${}_{2}$O${}_{8}$.
As in the previous figure, solid and dashed lines correspond to
simulations based on temperature independent and temperature dependent
homogeneous broadening, respectively.}
\label{XRef-Figure-523193659}
\end{figure}

We have already reported on {}\textsuperscript{51}V ($I=7/2$) NMR
measurements in the pristine PbNi${}_{2}$V${}_{2}$O${}_{8}$ compound
as well as in its Mg-substituted derivatives.\cite{ArconEPL65} The
temperature evolution of the NMR spectra for the latter case is
shown in Fig. \ref{XRef-Figure-523193630}(a) and Fig. \ref{XRef-Figure-523193659}(a).
We focus our attention on the broad spectra. A sharp resonance at
71.72~MHz is due to the presence of {}\textsuperscript{63}Cu nuclei
at or around our probe. Similarly, the additional narrow component
at around 71.42~MHz, observed in both Mg-doped samples will also
be neglected in the following analysis, as it shows no indications
of coupling of the detected nuclei to the electron spin system.\cite{ArconEPL65}

At room temperature the NMR absorption spectra of both Mg-doped
samples look qualitatively the same as those of the pristine compound.\cite{ArconEPL65}
They can be explained by the Hamiltonian
\begin{equation}
\mathcal{H}=\mathcal{H}_{Z}+\mathcal{H}_{Q}+\mathcal{H}_{hf}+\mathcal{H}_{cs},%
\label{XRef-Equation-526114458}
\end{equation}

\noindent where the first term corresponds to the Zeeman energy
of {}\textsuperscript{51}V nuclear spins {\bfseries I}${}_{\mathit{i}}$
in the external magnetic field {\bfseries B}${}_{0}$, the second
term $\mathcal{H}_{Q}$ represents the quadrupolar coupling of nuclear
quadrupolar moments with electric field gradients, the hyperfine
term $\mathcal{H}_{hf}$ originates from the transferred hyperfine
coupling of vanadium nuclear spins with surrounding nickel electron
spins {\bfseries S}${}_{\mathit{j}}$, and $\mathcal{H}_{cs}$ corresponds
to the chemical-shift Hamiltonian. The broad structure of the spectra
observed at room temperature originates from the quadrupolar coupling
with the consecutive satellite transitions displaced by $\nu _{Q}\approx
80$~kHz.\cite{ZorkoThesis} On the other hand, a pronounced broadening
of the central transition $(-1/2\rightarrow 1/2)$ is due to the
anisotropy of the chemical shift tensor.\cite{ZorkoThesis} Further,
the isotropic part of the transferred hyperfine interaction plays
the dominant role in determining the NMR shift. This fact is most
clearly manifested in a gap-like behavior of this parameter, which
thus scales with the bulk uniform susceptibility of the parent compound.\cite{ZorkoThesis}

To evaluate the {}\textsuperscript{51}V NMR frequency shift the
standard VOCl${}_{3}$ compound was utilized. Its {}\textsuperscript{51}V
NMR spectrum is positioned at $\nu _{L}^{dia}=70.974$~MHz in our
magnetic field. Since the measured frequency shift in the PbNi${}_{2}$V${}_{2}$O${}_{8}$
sample is a combined contribution of the isotropic part of the transferred
hyperfine interaction and the isotropic part of the chemical shift
tensor, the Larmor frequency of exactly $\nu _{L}^{0}=71.0$~MHz
corresponding to the pristine compound at 4.2~K, can be used for
setting the reference of a zero paramagnetic shift. Namely, at such
low temperatures the spin susceptibility on the Haldane spin system
should be diminished due to its activated behavior. The value $\nu
_{L}^{0}$ is presented in our figures with NMR spectra by horizontal
lines.
\begin{figure}[ht]
\begin{center}
\includegraphics[width=7.2cm]{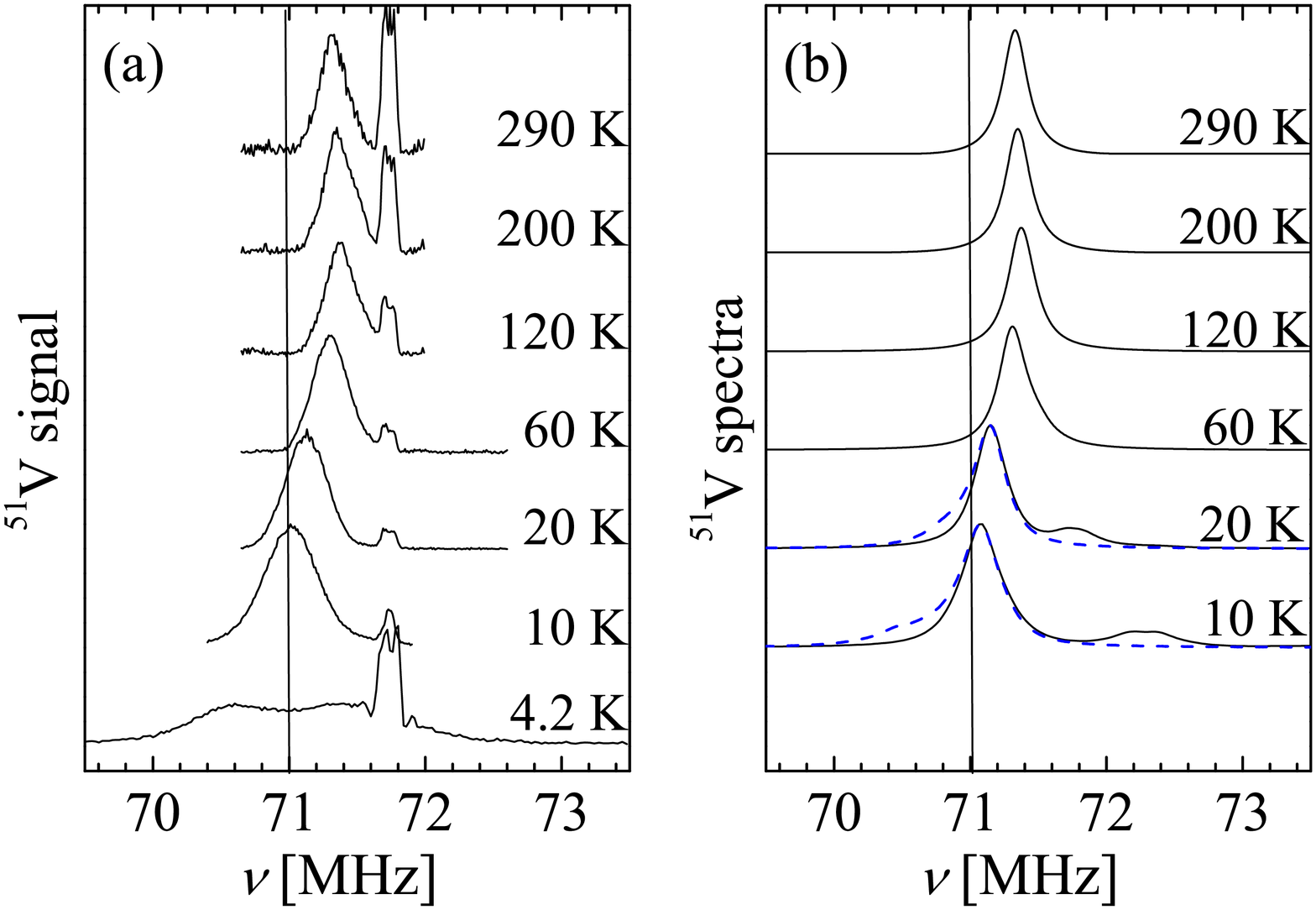}

\end{center}
\caption{ The temperature evolution of (a) measured and (b) simulated
{}\textsuperscript{51}V NMR spectra in PbNi${}_{1.92}$Co${}_{0.08}$V${}_{2}$O${}_{8}$.
The fits correspond to paramagnetic order (solid lines) and antiferromagnetic
(dashed lines) correlations between impurity and impurity-induced
spins.}
\label{XRef-Figure-523215933}
\end{figure}

When lowering the temperature, the spectra of the Mg-doped compounds
exhibits a pronounced broadening with respect to the parent compound,
which shows practically unchanged linewidth in the whole investigated
temperature range. We have previously speculated that this experimental
finding should be a precursor effect of the three-dimensional magnetic
ordering in both Mg-doped compounds.\cite{ArconEPL65} In this paper,
we offer a quantitative description of the broadening effect, which
is based on the temperature evolution of the size of the impurity-induced
staggered moments as well as on the development of the three-dimensional
correlations between them.

Contrary to the case of the nonmagnetic Mg${}^{2+}$ doping, magnetic
Co${}^{2+}$ dopants have a completely different impact on the evolution
of the {}\textsuperscript{51}V NMR spectra. As shown in Fig \ref{XRef-Figure-523215933}(a),
in PbNi${}_{1.92}$Co${}_{0.08}$V${}_{2}$O${}_{8}$ the spectra show
less pronounced broadening with decreasing temperature. They, however,
exhibit a drastic change in their width and appearance (two-peak
structure) below 6~K, signaling the transition into the magnetically
ordered state. The broadening effect is even more reduced in the
case of the PbNi${}_{1.98}$Co${}_{0.02}$V${}_{2}$O${}_{8}$ compound
as presented in the next section. The observation of broader NMR
spectra in nonmagnetically doped compounds than in magnetically
doped samples seems surprising since magnetic impurities are expected
to be coupled to the vanadium nuclei in contrast to the vacant magnesium
sites. Thus, at least close to the phase-transition temperature
Co${}^{2+}$ impurities should significantly contribute to the appearance
of the NMR spectra due to slowing-down of the electron spin fluctuations.
\begin{figure}[ht]
\begin{center}
\includegraphics[width=7.2cm]{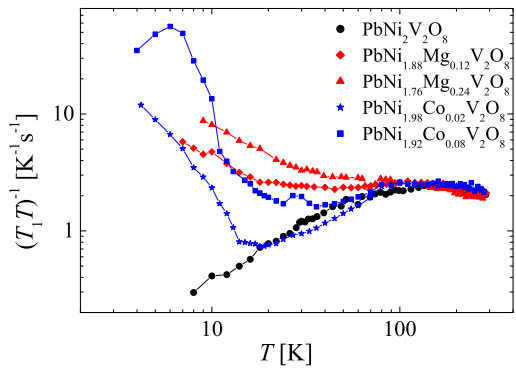}

\end{center}
\caption{ (Color online) The temperature dependence of the {}\textsuperscript{51}V
NMR spin-lattice relaxation rate (divided by temperature) in the
parent and the impurity doped PbNi${}_{2}$V${}_{2}$O${}_{8}$ compounds.}
\label{XRef-Figure-526145111}
\end{figure}

To get a further insight into the temperature development of the
electron spin correlations, we also performed {}\textsuperscript{51}V
NMR spin-lattice relaxation measurements. Due to strong transferred
hyperfine coupling the spin-lattice relaxation is expected to be
mainly caused by transverse electron spin fluctuations.\cite{MoriyaPTP16,MoriyaPTP28}
When the temperature is lowered a rather diverse behavior is observed
in different samples. The monotonic decrease of the parameter ${(T_{1}T)}^{-1}$
with decreasing temperature as observed in the parent compound,
is characteristic of the Haldane-gap excitations.\cite{SagiPRB53}
On the other hand, the low-temperature up-turn in all doped samples
gives a clear indication of the developing electron spin correlations,
as discussed further in the next section. A rather surprising observation
is again the fact that the low-temperature deviations of the ${(T_{1}T)}^{-1}$
in Mg-doped samples with respect to the pristine compound are observed
from much higher temperatures than in Co-doped samples, while at
the same time the increase of this parameter is much more moderate
in the case of the nonmagnetic impurities. Similarly to the bulk
magnetic measurement and the temperature evolution of the NMR spectra,
these experimental findings again indicate on the impurity dependent
character of the electron spin correlations.

\section{ANALYSIS AND DISCUSSION}

\subsection{Frequency shift and broadening of the {}\textsuperscript{51}V
NMR spectra in doped compounds}

As already mentioned in the preceding section, the room-temperature
lineshape of the {}\textsuperscript{51}V NMR spectra in the parent
as well as in all the doped samples is determined by the quadrupolar
Hamiltonian and the anisotropic part of the chemical shift tensor.
In addition, the relatively large shift of the spectra, i.e., $\Delta
\nu /\nu _{L}^{0}=0.4\%$, can be attributed to the transferred hyperfine
coupling between the vanadium nuclei and the electron magnetic moments
localized on Ni${}^{2+}$ (and impurity) sites,
\begin{equation}
\mathcal{H}_{hf}=\sum \limits_{i,j}{\textbf{I}}_{i}\cdot {\textbf{A}}_{i,j}\cdot
{\textbf{S}}_{j},%
\label{XRef-Equation-52611455}
\end{equation}

\noindent where the sums run over all the vanadium nuclei ({\bfseries
I}${}_{\mathit{i}}$) and over its six nearest-neighbor Ni${}^{2+}$
sites ({\bfseries S}${}_{\mathit{j}}$).\cite{MastorakiJSSC177,ZorkoThesis}
Although isolated V${}^{5+}$ ions are diamagnetic, they can be addressed
as being partially magnetic\cite{JaccarinoAFM} in the PbNi${}_{2}$V${}_{2}$O${}_{8}$
compound. This is due to the imbalance in the closed vanadium electronic
shells caused by the interaction of vanadium and nickel electrons,
which results in the effective transferred hyperfine coupling between
the {}\textsuperscript{51}V nuclear spins and the Ni${}^{2+}$ electron
spins. In powder samples only the isotropic part of the hyperfine
tensor contributes to the shift of the NMR spectra, which allows
us to estimate this coupling as 
\begin{equation}
A_{i,j}^{iso}=\frac{2h \Delta \nu  N_{A}g \mu _{B}}{6\chi _{mol}B_{0}}=0.17~\mathrm{mK}.%
\label{XRef-Equation-525153212}
\end{equation}

\noindent In the above estimation we used the room-temperature value
of the molar susceptibility in the magnetic field of 5~T, $\chi
_{mol}= 5.6\cdot {10}^{-3}$ emu/mol, as a measure of the average
value of the nickel spins in the parent material $\langle S_{j}^{0}\rangle
=\chi _{mol}B_{0}/2N_{A}g \mu _{B}$. Room-temperature {\itshape
g}-factor value $g=2.2$ was determined by ESR measurement.\cite{ZorkoPRB65}

In doped Haldane chains impurity-induced staggered magnetic spins
$\langle S_{j}^{i}\rangle $ are expected to be superimposed on the
uniform spin chain, defining the average spin value at each site
as
\begin{equation}
\left\langle  S_{j}\right\rangle  = \left\langle  S_{j}^{0}\right\rangle
+\left\langle  S_{j}^{i}\right\rangle  .%
\label{XRef-Equation-5251642}
\end{equation}

\noindent An NMR detection of the staggered magnetization near impurity
sites has been, in fact, recently reported in another Haldane chain
compound Y${}_{2}$BaNiO${}_{5}$.\cite{TedoldiPRL83} The authors
observed additional peaks shifted to lower and higher frequencies
with respect to the main NMR line of the {}\textsuperscript{89}Y
nuclei being coupled to single Ni${}^{2+}$ sites. These lines were
attributed to the yttrium nuclei coupled to the staggered moments,
with exponentially decaying correlations. The temperature dependence
of the correlation length was shown to follow nicely theoretical
predictions,\cite{KimEPJB4} while the amplitude of the staggered
moments exhibited Curie-dependence for $S=1/2$ spins, in the case
of nonmagnetic Zn${}^{2+}$ as well as magnetic Cu${}^{2+}$ dopants.\cite{DasPRB69}
Well defined chain-end excitations, experimentally observed also
at temperatures far above the Haldane gap, where theoretically reproduced
by quantum Monte Carlo calculations,\cite{AletPRB62} which justified
the simple decomposition of the average spin value given by Eq.
(\ref{XRef-Equation-5251642}) in a broad temperature range.

\begin{figure}[t]
\begin{center}
\includegraphics[width=6.0cm]{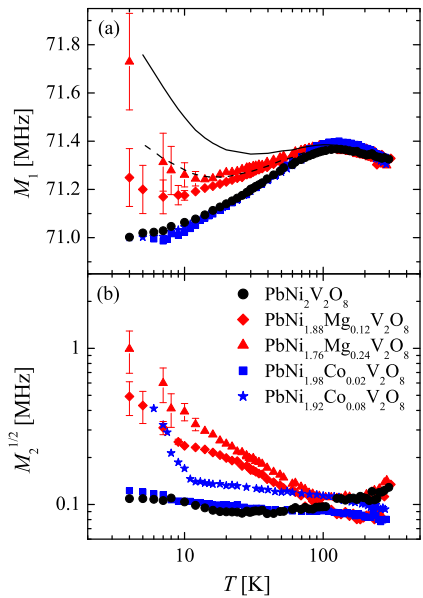}

\end{center}
\caption{ (Color online) The temperature dependence of the {}\textsuperscript{51}V
NMR (a) line position and (b) linewidth in doped PbNi${}_{2}$V${}_{2}$O${}_{8}$
compounds. The lines correspond to the model explained in the Section
\ref{XRef-Section-629155821}.}
\label{XRef-Figure-525131740}
\end{figure}

The shift of the NMR spectra in doped PbNi${}_{2}$V${}_{2}$O${}_{8}$
compounds should reflect the impurity-induced contribution to the
dc magnetization of these compounds. In Fig. \ref{XRef-Figure-525131740}(a)
the temperature dependence of the first moment of the measured {}\textsuperscript{51}V
NMR spectra is presented. In the case of the nonmagnetic Mg-doping
the expected up-turn in the NMR shift is observed at low temperatures,
however, the increase of the low-temperature dc susceptibility is
much more pronounced (see Fig. \ref{XRef-Figure-523213316}). Disagreement
between the two quantities is even more obvious in the case of the
magnetic Co-doping. Here, the impurity contribution to the magnetization
observed in the dc susceptibility at low-temperatures is not reflected
in the shift of the NMR spectra at all. Although, at first sight,
these observations would suggest that our NMR measurements can not
detect the impurity-induced staggered magnetization, we prove in
the subsequent subsection that this is not the case. Namely, the
fact that vacant sites (Mg${}^{2+}$ impurities) produce shifts towards
lower frequencies and that Co${}^{2+}$ impurities induce weaker
hyperfine coupling to {}\textsuperscript{51}V nuclei, prove to be
essential.

The linewidth (second moment) of the NMR spectra also significantly
depends on the spin nature of the dopants [Fig. \ref{XRef-Figure-525131740}(b)].
The broadening with lowering temperature for the Mg-doped samples is tremendous and starts
at rather high temperatures with respect to the phase-transition
temperature in low magnetic fields. On the other hand, the Co-doped
samples show much more moderate broadening 
in the paramagnetic phase and a significant increase of the linewidth
in the vicinity of the phase transition, characteristic of critical
effects, as explained below. This fact implies on a diverse nature
of the spin correlations determining the low-temperature NMR lineshape
in both cases, which is in line with the drastically different stability
of the magnetically ordered phase.

\subsection{Simulation of the {}\textsuperscript{51}V NMR spectra}\label{XRef-Section-629155821}

Contrary to the {}\textsuperscript{89}Y ($I=1/2$) case in Y${}_{2}$BaNiO${}_{5}$,\cite{TedoldiPRL83,DasPRB69}
the NMR spectra of the {}\textsuperscript{51}V nuclei in PbNi${}_{2}$V${}_{2}$O${}_{8}$
are quadrupolarly broadened. For this reason, the impurity-induced
inhomogeneities of the spin density sensed by vanadium nuclei are
hidden within the spectra at high temperatures. For instance, at
room temperature a spin following the Curie dependence and corresponding
to the $S=1/2$ degree of freedom induces according to Eq. (\ref{XRef-Equation-525153212})
a frequency shift of ${\Delta \nu }^{i}=30$~kHz at the {}\textsuperscript{51}V
nucleus, to which it is coupled. This value should be compared to
the width of the room-temperature experimental spectra, approximately
given by $2\nu _{Q}=160$~kHz. However, at lower temperatures the
impurity-generated features of the NMR spectra are expected to come
into sight.

Trying to construct a quantitative picture of the above-mentioned
effects, we performed the following simulation procedure. In a finite
spin chain ($N=4096$) the impurity positions with a given concentration
were randomly distributed. The uniform part $\langle S_{j}^{0}\rangle
$ of the average spin value at a certain temperature was deduced
from the temperature dependence of the position of the {}\textsuperscript{51}V
NMR spectra in the pristine compound. The contribution of the staggered
moments, $\langle S_{j}^{i}\rangle $, was calculated from the position
of all impurities, taking into account the staggered nature of the
impurity-induced moments with the theoretically predicted temperature-dependent
correlation length\cite{KimEPJB4} and assuming the magnitude of
these moments to be given by the Brillouin function, $S\cdot B_{S}(
g \mu _{B}S B_{0}/k_{B}T) $. In addition, in the case of Co-doping,
the spin of the impurities $S_{i}=3/2$, following the Curie dependence,
was also taken into consideration. The average spin value interacting
with a particular {}\textsuperscript{51}V nucleus was then calculated
as a sum of four consecutive spins on Ni${}^{2+}$ sites in one chain
and two spins in its neighboring chain. Namely, each vanadium nucleus
is coupled through an oxygen bridge to four {\itshape nn} nickel
ions in one chain and to two {\itshape nn} ions in the neighboring
chain. As the vanadium-nickel distances and the bridging angles
are very similar,\cite{MastorakiJSSC177} it is reasonable to first
assume a common value of the isotropic hyperfine coupling, evaluated
in Eq. (\ref{XRef-Equation-525153212}). On the contrary, a reduced
hyperfine coupling was considered for Co${}^{2+}$ impurities, ${(A_{\mbox{}ij}^{iso})}^{i}=A_{ij}^{iso}J_{i-h}/J$,
in accordance with the reduction of the impurity-host exchange {\itshape
J}${}_{i-h}$ with respect to the {\itshape nn} intrachain exchange
{\itshape J}.\cite{ZorkoESRCoMg} From the distributions of the average
electron spin values sensed by {}\textsuperscript{51}V nuclei, ``stick''
diagrams of the NMR spectra could be obtained. These diagrams were
then convoluted by a homogeneously broadened Lorentzian line with
the linewidth corresponding to the room-temperature experimental
spectra. 

\subsubsection{Mg-doping}

The temperature evolution of the simulated spectra for PbNi${}_{1.88}$Mg${}_{0.12}$V${}_{2}$O${}_{8}$
and PbNi${}_{1.76}$Mg${}_{0.24}$V${}_{2}$O${}_{8}$ is presented
in Fig. \ref{XRef-Figure-523193630}(b) and Fig. \ref{XRef-Figure-523193659}(b),
respectively. The solid lines are calculated on the assumption of
temperature independent homogeneous broadening, which corresponds to the case
of the parent compound, where the lineshape is given by the quadrupole
Hamiltonian. Although the exact agreement with the experimental
spectra at lower temperatur es is not reached, the impurity-induced
shift of a significant portion of the signal towards higher frequencies
is in a nice qualitative agreement with the asymmetric broadening
of the NMR spectra observed in both Mg-doped samples. When a temperature
dependent homogeneous broadening of the lines is assumed, the experimental
NMR spectra are adequately reproduced, as shown by the dashed lines
in Fig. \ref{XRef-Figure-523193630}(b) and Fig. \ref{XRef-Figure-523193659}(b).
The fits were made down to 10~K because at lower temperatures the
assumed Curie dependence should become inappropriate due to the
proximity with the antiferromagnetic phase region. The predicted
spectrum at 10~K is calculated with 3-times larger homogeneous broadening
than the spectrum at room temperature. The necessity of including
enhanced homogeneous broadening at temperatures significantly above
the ordering temperature in low magnetic fields reopens our initial
assumptions that the three-dimensional ordering effects resulting
in the broadening of the NMR spectra, should be important already
at rather high temperatures.\cite{ArconEPL65}

This model can be also used for the prediction of the NMR frequency shift.
The calculated temperature-dependence is given in Fig. \ref{XRef-Figure-525131740}(a)
by a dashed and a solid lines, for the cases of PbNi${}_{1.88}$Mg${}_{0.12}$V${}_{2}$O${}_{8}$
and PbNi${}_{1.76}$Mg${}_{0.24}$V${}_{2}$O${}_{8}$, respectively.
At lower temperatures approximately 1.5-times larger shifts are
predicted than measured. This can be understood as a consequence
of the simplicity of our model taking into consideration the Curie
dependence of the staggered moments. Such assumption is reasonable
in the paramagnetic phase, yet close to the phase-transition point
a pronounced reduction of the moments is expected. Moreover, also
the correlation length should be affected by the vicinity of the
magnetically ordered phase. The increased correlation length produces
less inhomogeneous distribution of the spins sensed by the vanadium
nuclei and thus lower NMR shifts.

\subsubsection{Co-doping}

The calculated spectra in the case of the PbNi${}_{1.92}$Co${}_{0.08}$V${}_{2}$O${}_{8}$
compound are presented in Fig. \ref{XRef-Figure-523215933}(b). At
temperatures $T\gg J_{i-h}$ (14~K) a paramagnetic order of each
impurity spin and both impurity-induced staggered spins is expected,
which corresponds to the fits given by the solid lines. At the other
temperature limit, $T \ll J_{i-h}$, antiferromagnetic correlations,
producing effective spin $\tilde{S}=1/2$ at impurity sites are anticipated.
The simulated spectra with such correlations are presented by dashed
lines and agree reasonably with the measurements. The simulations
show slight broadening at lower temperatures because of the impurity-generated
inhomogeneities in the electron spin distribution. However, experimental
spectra suggest that also the homogeneous broadening must be enhanced
below approximately 20~K, which can be understood as a precursor
effect due to the proximity with the antiferromagnetic phase transition.

The phase transition into the magnetically ordered state occurs
at approximately 6~K in the magnetic field of 6.34~T. Below this
temperature the NMR spectra exhibit strong broadening and the lineshape
turns into a symmetric two-peak structure\cite{ZorkoThesis} as shown
in Fig. \ref{XRef-Figure-523215933}(a). Such lineshape can be regarded
as a signature of the bipartite spin lattice corresponding to the
antiferromagnetic order. However, as vanadium nuclei sense averaged
electron spin value over several sites, nonequivalent Ni-V bonds
are required. Different values of the coupling are plausible since
the Ni-V distances range from 3.33 {\AA} to 3.48 {\AA} and the Ni-O-V
bridging angles are spanned from 123$ \mbox{}^{\circ}$ to 135$ \mbox{}^{\circ}$
in PbNi${}_{1.88}$Mg${}_{0.12}$V${}_{2}$O${}_{8}$,\cite{MastorakiJSSC177}
which has a similar crystal structure to the PbNi${}_{1.92}$Co${}_{0.08}$V${}_{2}$O${}_{8}$
compound.

As seen from the simulated NMR spectra assuming antiferromagnetic
ordering of each Co${}^{2+}$ impurity spin and the two staggered
spins liberated next to it (Fig. \ref{XRef-Figure-523215933}), the
observed low-temperature discrepancy between the {}\textsuperscript{51}V
NMR line position shown in Fig. \ref{XRef-Figure-525131740}(a) and
the corresponding dc susceptibility of the PbNi${}_{1.92}$Co${}_{0.08}$V${}_{2}$O${}_{8}$
sample (Fig. \ref{XRef-Figure-523213316}) can be satisfactory explained.
It is due to the combined effect of the antiferromagnetic correlations
and the reduced transferred hyperfine coupling of the vanadium nuclei
with cobalt spins with respect to coupling to nickel spins. Even
more, the predicted NMR shifts at low temperatures displace the
lines slightly below the value of $\nu _{L}^{0}=71.0$~MHz. Such
negative shifts give a firm evidence for the validity of the presented
model.

The simulated spectra assuming antiferromagnetic correlations 
between each impurity spin and the two liberated host spins also
provide an adequate explanation why the inhomogeneous broadening of the NMR spectra
in Co-doped samples is severely reduced with respect to the case
of Mg-doping. The reason for the observed difference lies 
in the combined effect of the antiferromagnetic order and the 
reduced value of the cobalt-vanadium transferred hyperfine
exchange. Next, there seems to exist a difference also in the intrinsic
homogeneous broadening required to reproduce the low temperature
spectra. Surprisingly, this parameter is larger in the case of the
nonmagnetic doping, at least not in the extreme vicinity of the
phase-transition temperature. This unexpected observation can be
further inspected by analyzing the NMR spin-lattice relaxation,
which similarly displays the temperature evolution of the electron
spin correlations.

\subsection{Diverse nature of the electron spin correlations as
detected by the {}\textsuperscript{51}V spin-lattice relaxation}

The nuclear spin-lattice relaxation due to the hyperfine interaction,
given by Eq. (\ref{XRef-Equation-52611455}), is related to the transverse
components of the electron spin fluctuation $\delta \textbf{S}=$\textbf{S}-{\textlangle}\textbf{S}{\textrangle}.\cite{MoriyaPTP16,MoriyaPTP28}
In the case of the dominant isotropic hyperfine coupling the spin-lattice
relaxation rate is determined by the transverse spin correlations
$\langle {\delta S}_{\textbf{q}}^{+}( \tau ) {\delta S}_{-\textbf{q}}^{-}(
0) \rangle $, with {\bfseries q} being a wave vector. In the high-temperature
approximation the spin-lattice relaxation is given by\cite{MoriyaJPSJ18}
\begin{equation}
\frac{1}{T_{1}T}=\frac{k_{B}}{2\hbar ^{2}}\sum \limits_{\textbf{q}}A_{\textbf{q}}^{iso}A_{-\textbf{q}}^{iso}\frac{\chi
^{{\prime\prime}}( \textbf{q},\omega _{L}) }{\omega _{L}},
\end{equation}

\noindent where $A_{\textbf{q}}^{iso}$ is the Fourier transform
of the isotropic hyperfine coupling and $\chi ^{{\prime\prime}}(
\textbf{q},\omega _{L}) $ represents the dissipative part of the
transverse dynamical susceptibility.

When approaching the transition into the magnetically ordered state
the spin-lattice relaxation is expected to exhibit critical dependence,
due both to the divergent character of the static {\itshape q}-dependent
susceptibility in the center of the antiferromagnetic zone, as well
as due to the critical slowing down of the spin fluctuations.\cite{MoriyaPTP28}
It should be noted that the low-temperature increase of the spin-lattice
relaxation is much more enhanced than the increase of the static
uniform susceptibility reflected in the measurements of the dc susceptibility
and the NMR shift, because the spin-lattice relaxation detects spin
fluctuations at all wave vectors.

In Co-doped samples the above-mentioned expected behavior of the
spin-lattice relaxation is shown in Fig. \ref{XRef-Figure-526145111}.
The relaxation rate shows a strong temperature dependence close
to the phase-transition temperature and the peak observed for PbNi${}_{1.92}$Co${}_{0.08}$V${}_{2}$O${}_{8}$
at 6~K nicely corresponds with the occurrence of the two-peak structure
of the NMR spectra. On the contrary, the deviations of the spin-lattice
relaxation in Mg-doped samples from the dependence observed in the
pristine compound extend to much higher temperatures. At the same
time, however, they are suppressed with respect to Co-doping at
low temperatures. This observation suggests that the antiferromagnetic
correlations are inhibited in Mg-doped samples by the magnetic field,
which is in line with the bulk magnetization results in the magnetic
field of 5~T (Fig. \ref{XRef-Figure-523213316}). Therefore, enhanced
ferromagnetic correlations (reflected in increased $\chi ^{{\prime\prime}}(
0,\omega _{L}) $) should most likely be employed to account for
both the spin-lattice relaxation behavior and the observed homogeneous
broadening of the NMR spectra in the case of the nonmagnetic doping.
The NMR broadening detects also longitudinal spin correlations in
addition to the transversal correlations reflected in the spin-lattice
relaxation.\cite{MoriyaPTP28}

\subsection{Bulk magnetic properties}

As experimentally verified by both the bulk magnetization and the
NMR measurements, the magnetically ordered phase turns out to be
much more robust against the magnetic field in the case of the magnetic
Co-doping than in the case of the nonmagnetic Mg-doping. The metamagnetic
transition occurring around 1.4~T in the PbNi${}_{1.88}$Mg${}_{0.12}$V${}_{2}$O${}_{8}$
at 2~K is a consequence of a strong easy-axis single-ion anisotropy
{\itshape D} at each Ni${}^{2+}$ site,\cite{LappasPRB66} which is
in accord with the inelastic neutron scattering measurements, suggesting
$D=-5.2$~K.\cite{ZheludevPRB62} 

In general, phase-diagrams of uniaxial Heisenberg antiferromagnets
show three distinct phases depending on the value of the single-ion
anisotropy and the temperature; the paramagnetic phase, the antiferromagnetic
phase and the spin-flop phase between them. In the mean-field approximation
of the exchange interaction, it was theoretically predicted for
$S=1$ spin systems that a direct metamagnetic transition from the
antiferromagnetic to the paramagnetic phase occurs in magnetic fields
below $B_{c}=z J_{0}/g \mu _{B}$, if $D>z J_{0}$ ({\itshape z} is
the coordination number and {\itshape J}${}_{0}$ is the exchange
constant), with the critical field decreasing with temperature.\cite{VilfanJPC12}
In the present case of the one-dimensional spin system, which is
in addition highly inhomogeneous, the part of the exchange coupling
constant $z J_{0}$ is taken by the effective coupling providing
the three-dimensional magnetic ordering. The single-ion anisotropy
value $D=-5.2$~K should thus be compared to the phase-transition
temperature $T_{N}=3.4$~K. The zero-temperature value of the critical
field is then expected to be of the order $B_{c}\approx k_{B}T_{N}/g
\mu _{B}=2.7$~T. The magnetic field above the critical value breaks
the three-dimensional magnetic correlations between staggered moments.
However, due to the rather low value of the observed critical field,
the antiferromagnetic spin correlations within the individual staggered
moments should not be affected as they are governed by the strong
{\itshape nn} intrachain exchange. Such behavior was very recently
proposed to take place in lightly Mg-doped CuGeO${}_{3}$ compounds,
where a field-controlled microscopic separation of the antiferromagnetic
and the paramagnetic phase was achieved.\cite{GlazkovPRL94}

On the other hand, the magnetic ordering in the Co-doped samples
is preserved even in magnetic fields above 6~T as concluded from
our NMR measurement. At low doping concentrations of either magnetic
or nonmagnetic impurities the indirect staggered exchange\cite{SorensenPRB49}
$J_{s}( L) ={(-1)}^{L}0.81J \exp [ -(L-1)/\xi ] $, mediated by the
gapped Haldane medium, plays the dominant role in determining the
stability of the magnetic order. This prediction is confirmed by
the strong impurity-concentration dependence of the phase-transition
temperatures at low doping levels in both cases.\cite{ImaiCM2004}
However, at larger doping levels the cross-impurity exchange and
the interchain exchange between staggered moments, as weaker exchange
coupling mechanisms compared to the indirect staggered exchange,
determine the stability of the magnetic order. The observation of
the improved stability of the magnetically order phase in Co-doped
samples can then be attributed to the significantly enhanced three-dimensional
coupling of the impurity-induced staggered moments. The Co${}^{2+}$
impurities provide strong coupling between the two neighboring liberated
spin $S=1/2$ degrees of freedom due to the strong impurity-host
exchange, as well as a firmer interchain connections. In addition,
due to the strong spin-orbit coupling with respect to the crystal-field
splitting for Co${}^{2+}$ ions,\cite{PilbrowCo2+} an anisotropic
exchange is generally observed for these ions. Such anisotropy,
which reduces the spin fluctuations and thus enhances the stability
of the magnetically ordered state, should prove highly important
in the case of the Co-doping. Namely, doping with Co${}^{2+}$ ions
results in by far highest phase-transition temperatures with respect
to other magnetic impurities.\cite{ImaiCM2004}

Although, conceptionally, the impurity-induced long-range ordering
in different spin-gap systems should have a common origin, the system
dependent magnetic properties determining the stability of the ordered
state must not be neglected in future theoretical studies. For instance,
as extensive work on doping the spin-Peierls CuGeO${}_{3}$ compound
has shown, the spin of the dopants has only a minor influence on
the phase-transition temperature and the spin-flop behavior of the
antiferromagnetically ordered state. On the contrary, we have shown
that the stability of the magnetically ordered state in the PbNi${}_{2}$V${}_{2}$O${}_{8}$
Haldane compound crucially depends on the magnetic nature of the
impurities.

\section{CONCLUSIONS}

We have presented a study of the impurity-induced magnetic ordering
in the Haldane chain compound PbNi${}_{2}$V${}_{2}$O${}_{8}$. An
intrinsic difference between the effects of the nonmagnetic Mg${}^{2+}$
and the magnetic Co${}^{2+}$ impurities was observed. The long-range
magnetic order was shown to be destroyed in Mg-doped samples at
rather low magnetic fields implying that the antiferromagnetic correlations
within the impurity-liberated staggered magnetic moments should
stay preserved, as they are governed by the strong {\itshape nn}
intrachain exchange. On the other hand, the {}\textsuperscript{51}V
NMR as well as the bulk magnetization measurements gave a clear
indication of the improved stability of the antiferromagnetic order
in the Co-doped samples. This experimental finding was attributed
to the enhanced three-dimensional magnetic interactions between
localized staggered moments, provided by the substantial anisotropic
exchange coupling between impurity and host spins.

The staggered nature of the liberated degrees of freedom was indirectly
explored through the appearance of the {}\textsuperscript{51}V NMR
spectra. It was shown, that the broadening of the spectra as well
as the line shift correspond to the Curie dependence of the impurity-induced
staggered magnetic moments, delocalized on the scale of the theoretically
predicted correlation length. An expected deviation of the model
predictions from the experimental results was found only in the
vicinity of the phase transition. In addition, the NMR measurements
offered information about the different character of the electron spin correlations
for both types of doping. In Co-doped samples the critical behavior
of the NMR line broadening and the spin-lattice relaxation
was detected at temperatures close to the phase transition. On the
other hand, the pronounced homogeneous broadening of the NMR spectra
and the enhanced spin-lattice relaxation was observed in Mg-doped
samples at surprisingly high temperatures, despite the suppression
of the antiferromagnetic correlations upon the magnetic field, as
detected by the bulk magnetization measurements. This should be due 
to the enhanced spin correlations at $\textbf{q}=0$. It is envisaged
that this surprising behavior of the spin correlations in the nonmagnetically
doped samples can be further inspected by the complementary inelastic
neutron scattering (INS) measurements. As this experimental technique
provides both spectral and spacial information about spin fluctuations,
it should uncover additional important aspects of the magnetic-field
driven transformation, encompassing the nature of the spin fluctuations
during the metamagnetic transition in the doped Haldane-chain 
PbNi${}_{2}$V${}_{2}$O${}_{8}$ compound.

\acknowledgments We acknowledge the financial support provided by the General Secretariat
for Science \& Technology (Greece) and the former Ministry of Education,
Science and Sport of the Republic of Slovenia through a Greece-Slovenia
``Joint Research \& Technology Program''.

\end{document}